\documentstyle[12pt]{article}

\begin{document}

\title{Quantum Fields in Curved Spacetimes and Semiclassical
Approaches: A Workshop Summary} 
\author{Robert M. Wald\\ {\it Enrico Fermi Institute and
Department of Physics}\\ {\it University of Chicago}\\ {\it 5640
S. Ellis Avenue}\\ {\it Chicago, Illinois 60637-1433}} 

\maketitle

\begin{abstract}

I briefly review some of the recent progress in quantum field theory
in curved spacetime and other aspects of semiclassical gravity, as
reported at the D3 Workshop at GR15.
\end{abstract}

The study of the behavior of quantum fields in curved spacetime and
other semiclassical gravitational phenomena began in earnest in the
late 1960's with the study of particle creation effects in
cosmology. It then received enormous impetus in the mid 1970's from
Hawking's discovery of thermal particle creation effects by black
holes. In the late 1970's and early 1980's, a great deal of progress
was made in the development of the theory and the exploration of
various phenomena. By the 1990's, quantum field theory in curved
spacetime had become a mature subject, with its foundational
underpinnings well established.

In the past few years, some notable developments have occurred in a
number of aspects of quantum field theory in curved spacetime. These
developments include: (1) The application of the ``microlocal
analysis'' methods of Hormander to resolve a number of outstanding
issues, such as the definition of the stress-energy tensor of a free
field as an operator-valued-distribution \cite{bfk} and the
renormalizability of interacting quantum field perturbative expansions
\cite{bf}. (2) The exploration of issues concerning quantum field
theory in non-globally-hyperbolic spacetimes and the analysis of the
behavior of quantum fields near various types of Cauchy horizons. (3)
The derivation and analysis of some global energy inequalities
satisfied by the expected stress-energy tensor of a quantum field. At
the D3 workshop at GR15, there were no contributions presented on
developments directly related to (1), but developments related to (2)
were very well represented in the contributions by Flanagan, Barve,
Hiscock, Kay, and Higuchi, and some developments in area (3) were
reported in the contribution by Ford.

In addition to research on topics falling within the precise confines
of quantum field theory in a curved spacetime, there also was
considerable research activity during the past few years in other
areas of ``semiclassical gravity'', involving issues such as pair
creation of black holes and black hole thermodynamics. The
contributions presented at the D3 workshop by Frolov, Bousso, Spindel,
and Manogue provide some representation of a few of these other
developments in semiclassical gravity.

In the following, I will provide a ``thumbnail sketch'' of the 12 oral
contributions presented at the D3 workshop. In most cases, references
to papers giving a complete exposition of the work will be provided,
and the interested reader is strongly advised to consult those
references rather than relying on the ``sound bites'' provided
here. Unfortunately, it is not feasible for me to attempt to summarize
in any way the nearly 30 contributions accepted for poster
presentations at the D3 workshop.

There has been considerable interest in understanding the stability of
Cauchy horizons such as the ``inner horizon'' of the
Reissner-Nordstrom and Kerr black holes. The classical ``blueshift
instability'' of these horizons is now well understood. However, there
are examples known of spacetimes where such a classical blueshift
instability does not occur, but, nevertheless, the Cauchy horizon is
unstable when perturbed by a test quantum field. The contribution by
E. Flanagan considered two-dimensional spacetimes where the classical
blueshift instability does not occur. Flanagan obtained a necessary condition
for these Cauchy horizons to be semiclassically unstable. He also
showed that the quantum instability of these these horizons could be
interpreted as resulting from a ``delayed blueshift''
instability. Details of this work can be found in \cite{f}.

The contribution by S. Barve (reporting on work done in collaboration
with T.P. Singh, C. Vaz, and L. Witten) considered the issue of the
quantum instability of a Cauchy horizon occurring when a naked
singularity arises in a $(1+1)$-dimensional model of the Tolman-Bondi
collapse of a ball of dust. It was found that the quantum
stress-tensor diverges on the Cauchy horizon in a manner similar to
what was previously found to occur in the case of naked singularities
produced from the collapse of null dust \cite{hwe}. Details
can be found in \cite{bsvw}.

The contribution by W.A. Hiscock noted that the extreme
Reissner-Nordstrom solution could be the asymptotic final state of a
black hole if magnetic monopoles or other suitable $U(1)$ charges
exist in nature. The usual blueshift instability arguments would not
apply to the Cauchy horizon of a black hole which asymptotically
becomes extreme, since the surface gravity of the Cauchy horizon would
vanish. However, Hiscock considered a Reissner-Nordstrom-Vaidya model
of the approach to an asymptotically extreme black hole, and found
that in this model the Cauchy horizon is singular nevertheless.

A contribution by B.S. Kay (reporting on work done in collaboration
with A.R. Borrott) also considered the extreme Reissner-Nordstom
spacetime, but concerned itself with the behavior of the quantum field
on the black hole event horizon. It is known \cite{t} that in 2
spacetime dimensions, there is a ``weak divergence'' of stress-energy
tensor at the horizon in the vacuum state (and a ``strong divergence''
for all thermal states at finite temperatures). However, in 4
spacetime dimensions, no divergences occur if one approximates the
extreme Reissner-Nordstrom metric by a Robinson-Bertotti metric
\cite{m}. (Numerical work on the 4-dimensional Reissner-Nordstrom
spacetime itself also indicates the absence of divergences
\cite{ahl}.) However, Borrott and Kay have found that in 2 dimensions,
the behavior of a quantum field in the vacuum state near the horizon
of the extreme Reissner-Nordstrom black hole differs from its behavior in
the Robinson-Bertotti spacetime in that Hessling's ``quantum
equivalence principle'' fails for the former but holds for the latter.
This suggests that great caution must be exercised in using the
Robinson-Bertotti approximation to the extreme Reissner-Nordstrom
black hole.

The contribution of A. Higuchi (reporting on work done in
collaboration with C.J. Fewster and B.S. Kay) dealt with the
construction of quantum field theory in chronology violating---and,
thus, non-globally-hyperbolic---spacetimes. Kay has proposed the
condition of ``F-locality'' as a necessary criterion to be satisfied
by a quantum field theory in a non-globally-hyperbolic spacetime
\cite{k}. (This condition asserts that every point should have a
globally hyperbolic neighborhood ${\cal U}$ such that the restriction
of the field algebra to ${\cal U}$ agrees with what one would obtain
by viewing ${\cal U}$ as a globally hyperbolic spacetime in its own
right.) It is known that F-locality must fail in any spacetime with a
compactly generated chronology horizon \cite{krw}, but some examples
(like the ``spacelike cylinder'') are known of chronology violating
spacetimes which admit F-local field algebras. However, the recent
work reported by Higuchi on conformal deformations of the
4-dimensional spacelike cylinder provides some evidence that the
chronology violating spacetimes which admit F-local field algebras may
be non-generic.

The behavior of a quantum field near a chronology horizon was the
subject of a contribution by B.S. Kay reporting on work done with
C.R. Cramer. (This contribution was scheduled to be presented by
Cramer, but Cramer was unable to attend the meeting.) Although a
general theorem \cite{krw} establishes that the stress-tensor of a
quantum field must always be singular on a compactly generated
chronology horizon, explicit examples are known of states on 2- and
4-dimensional Misner spacetime for which the stress-energy tensor does
not diverge as one approaches the chronology horizon. These examples
were re-examined by Cramer and Kay using image sum techniques in order
to gain insight into how and why the singularity predicted by
\cite{krw} occurs nevertheless. Details of this work can be found in
\cite{ck}.

The contribution by L.H. Ford (reporting on work done in collaboration
with M.J. Pfenning) was concerned with restrictions on the energy
density of a quantum field in curved spacetime. It is well known that
none of the local (pointwise) energy conditions of classical field
theory apply to the expected stress-energy of a quantum field: For any
point $p$ in spacetime, one can find states that make the energy
density at $p$ be arbitrarily negative. However, in flat spacetime, a
number of global restrictions occur. In particular, there exist
``quantum inequalities'', which, roughly speaking, state that a
geodesic observer cannot observe a time averaged energy density more
negative than $- \hbar /t^4$, where $t$ denotes the ``sampling
time''. The main purpose of the present work by Ford and Pfenning was
to generalize the quantum inequalities to static curved spacetimes. It
was shown that for short sampling times, quantum inequalities exist
and take the form of the flat spacetime result plus subdominant,
spacetime-dependent corrections. Furthermore, they showed that the
average energy density measured along the worldline of a static
observer is bounded from below by the vacuum energy density. Details
of this work can be found in \cite{fp}.

Quantum field theory on a stationary spacetime containing an
``ergoregion'' but no black hole (as would occur for a solution
describing a sufficiently rapidly rotating relativistic star) was
considered in the contribution by G. Kang. Such spacetimes are known
to be classically unstable, and the presence of unstable modes
poses some difficulties for formulating the canonical quantization of
a scalar field if one tries to use the same procedures that are
applicable in stationary spacetimes without ergoregions. Kang
presented a prescription for quantizing a scalar field in the presence
of classically unstable modes, and showed how the ergoregion
instability persists in the quantum theory. Details of this work can
be found in \cite{kang}

V.P. Frolov reported on work he and his collaborators have done during
the past few years with the aim of accounting for the Bekenstein-Hawking
formula, $S = A/4$, for the entropy of a black hole in the context of
Sakharov's theory of induced gravity. In Sakharov's proposal, the
dynamical aspects of gravity arise from the collective excitations of
massive fields. Constraints are then placed on these massive fields to
cancel divergences and ensure that the effective cosmological constant
vanishes. In the induced gravity model explicitly considered by
Frolov, the Bekenstein-Hawking entropy is then explained as arising
from the ordinary statistical mechanical entropy of a thermally
excited gas of the heavy constituents. Details of this work can be
found in \cite{frolov}.

The contribution of R. Bousso (reporting on research done in
collaboration with S.W. Hawking) concerned the quantum behavior of
Schwarzschild black holes in de Sitter spacetime. A Schwarzschild
black hole in an asymptotically flat spacetime will evaporate via
Hawking radiation, and the same should be true for a ``small''
Schwarzschild black hole in de Sitter spacetime, since the temperature
of the black hole will be larger than that of the cosmological
horizon. However, when the black hole is of maximal mass
(corresponding to the Narai solution), the black hole and cosmological
temperatures are equal, so equilibrium should be possible. However,
one would expect this equilibrium to be unstable, since if the black
hole loses mass it should become hotter and, hence, should continue to
evaporate. Bousso and Hawking analyzed this issue and found that
stability actually occurs for a certain class of metric perturbations,
but that unstable modes exist for perturbations outside of this
class. Furthermore, they argued that these unstable modes should be
excited if black holes are produced by spontaneous pair
creation. Details of this work can be found in \cite{bh}.

P. Spindel reported on research (done in collaboration with
C. Gabriel, S. Massar, and R. Parentani) concerning the Unruh
effect. In Unruh's original model of an accelerating detector, the
internal states of the detector were treated quantum mechanically, but
the acceleration of the world line of the detector was treated as
being classically imposed. The work reported by Spindel provided a
fully quantum field theoretic model of an Unruh detector in which one
has charged fields of masses $M$ and $m$ which are placed in a uniform
electric field (which provides the acceleration) and these massive
fields are allowed to interact via coupling to yet another
field. Thermal effects are then investigated by comparing the ratio of
populations of the two species of massive particles. In the limit
where $M$ and $m$ go to infinity with $(M-m)$ and the acceleration, $a$,
fixed, and in the limit where the charge of the field which couples
the two massive fields goes to zero, the ratio of these populations
was found to correspond to a thermal distribution at the Unruh
temperature $a/2 \pi$. Details of this work can be found in
\cite{gsmp}.

The final oral presentation of the workshop was given by C.A. Manogue,
presenting work (done in collaboration with T. Dray) on a new
dimensional reduction scheme. In this approach, one starts with the
momentum space Dirac equation in 10 dimensions, described in terms of
2-component spinors over the octonions. The choice of a preferred
octonion unit is then used to effectively reduce the 10 spacetime
dimensions to 4 without resorting to compactification. The preferred
octonion unit also singles out 3 quaternionic subalgebras of the
octonions, which are interpreted as corresponding to the 3 generations
of leptons, whose massless particles naturally have just a single
helicity. Details of this work are given in \cite{md}.

\end{document}